\documentclass[aps,pre,amsmath,amssymb,twocolumn,showpacs]{revtex4}     
\usepackage{graphicx}
\usepackage{afterpage}

\begin{document}
\title{Spatial Correlation of the Dynamical Heterogeneity in a Binary Lennard-Jones Liquid in the Isoconfigurational Ensemble}
\title{Spatial Correlation of the Dynamic Propensity in a Glass-Forming Liquid}

\author{M. Shajahan G. Razul}
\affiliation{Department of Physics, St. Francis Xavier University,
Antigonish, NS, B2G 2W5, Canada}

\author{Gurpreet S. Matharoo}
\affiliation{Department of Physics, St. Francis Xavier University,
Antigonish, NS, B2G 2W5, Canada}

\author{Peter H. Poole}
\email{ppoole@stfx.ca}
\affiliation{Department of Physics, St. Francis Xavier University,
Antigonish, NS, B2G 2W5, Canada}

\begin{abstract}

We present computer simulation results on the dynamic propensity [as defined by Widmer-Cooper, Harrowell, and Fynewever,  Phys. Rev. Lett. {\bf 93}, 135701 (2004)] in a Kob-Andersen binary Lennard-Jones liquid system consisting of 8788 particles.  We compute the spatial correlation function of the dynamic propensity as a function of both the reduced temperature $T$, and the time scale on which the particle displacements are measured.  For $T\leq 0.6$, we find that non-zero correlations occur at the largest length scale accessible in our system.  We also show that a cluster-size analysis of particles with extremal values of the dynamic propensity, as well as 3D visualizations, reveal spatially correlated regions that approach the size of our system as $T$ decreases, consistent with the behavior of the spatial correlation function.  Next, we define and examine the ``coordination propensity", the isoconfigurational average of the coordination number of the minority B particles around the majority A particles.  We show that a significant correlation exists between the spatial fluctuations of the dynamic and coordination propensities.  In addition, we find non-zero correlations of the coordination propensity occurring at the largest length scale accessible in our system for all $T$ in the range $0.466<T<1.0$.  We discuss the implications of these results for understanding the length scales of dynamical heterogeneity in glass-forming liquids.

\end{abstract}


\date{\today}
\maketitle

\section{Introduction}

Glass-forming liquids are remarkable for the extraordinary sensitivity of their transport properties to changes in state variables, such as temperature $T$~\cite{pablo}.  Properties such as viscosity are commonly found to vary over 14 orders of magnitude in supercooled liquids between the melting temperature and the glass transition.  Yet in the same interval of $T$, it is also typical that only modest changes occur in the average liquid structure.  One of the central questions in the study of the glass transition is whether this enormous dynamical response can be understood in terms of structural change~\cite{DS01}.

\begin{figure}[t]
\centerline{\includegraphics[scale=0.35]{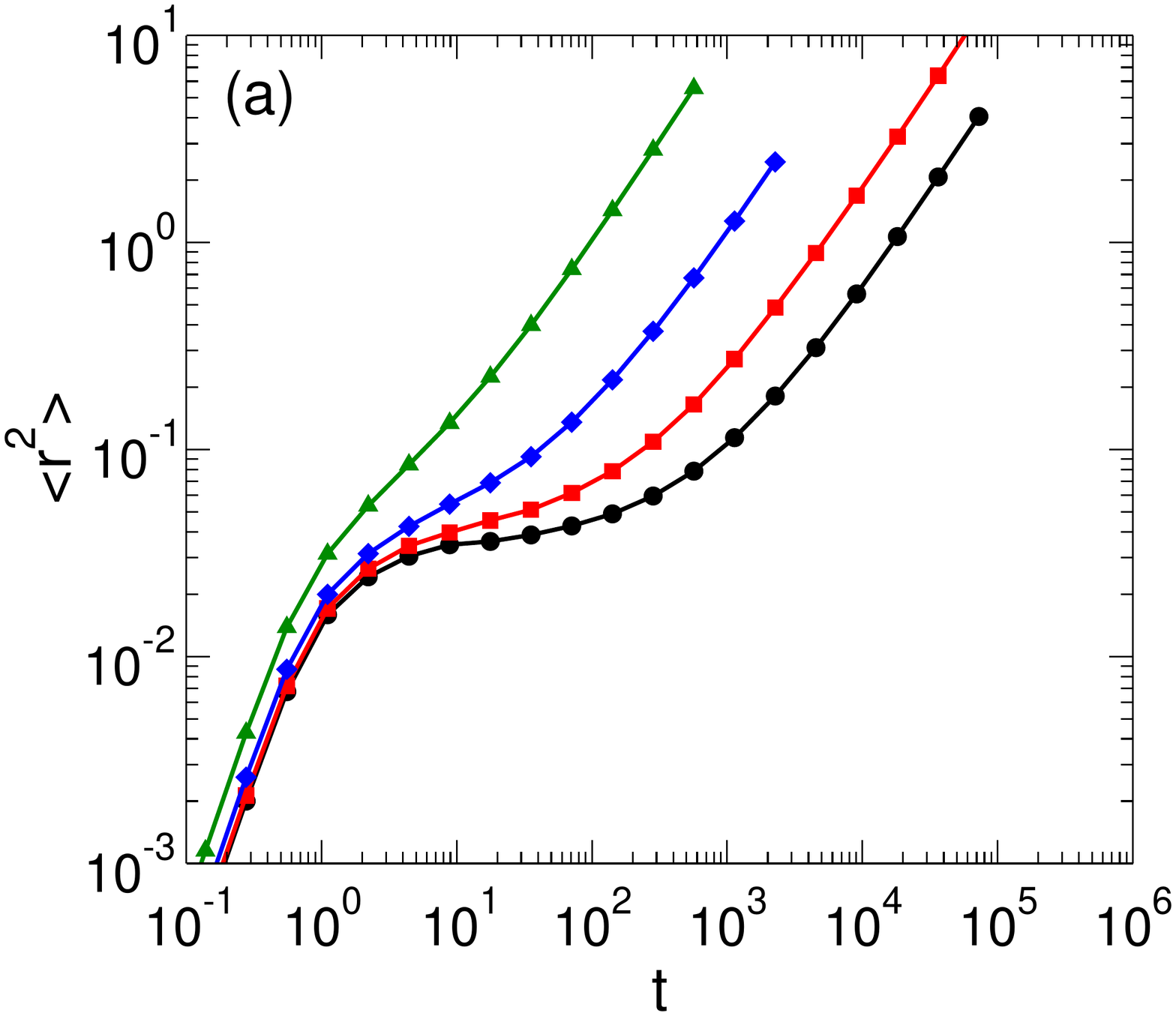}}
\bigskip
\bigskip
\centerline{\includegraphics[scale=0.35]{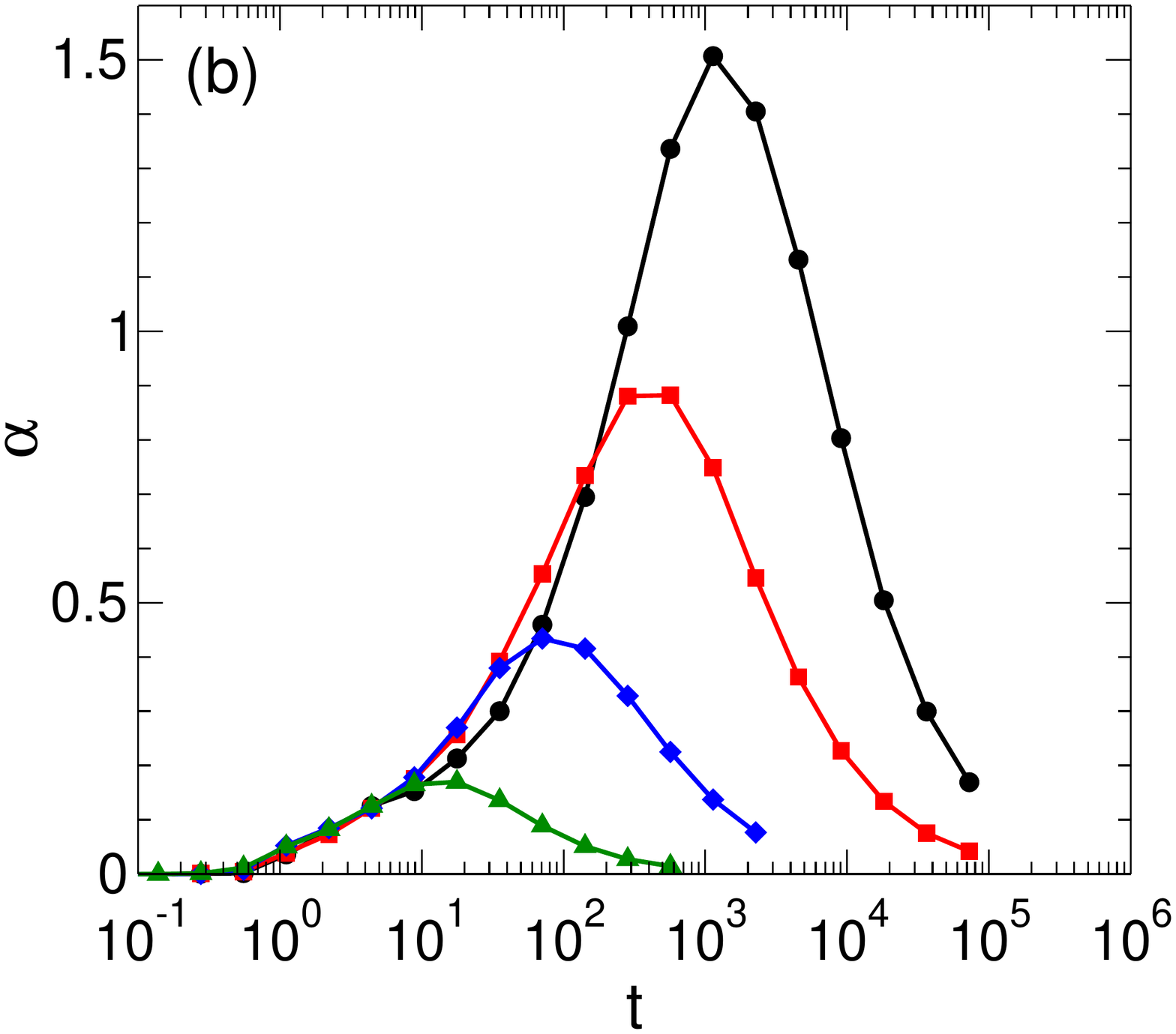}}
\caption{(a) Mean squared displacement $\langle r^2 \rangle$ and (b) non-Gaussian parameter $\alpha$ as a function of time, for $T=1.0$ (triangles), 0.6 (diamonds), 0.5 (squares), and 0.466 (circles).}
\label{ngp}
\end{figure}

Much recent interest has focussed on the emergence and growth of dynamical heterogeneity (DH) in supercooled liquids, that is, spatially extended domains in which molecules are more or less mobile, relative to the bulk average~\cite{A05}.  The growth of these dynamical domains as $T$ decreases seems to occur in the absence of a growing structural length scale.  At the same time, the simulation work of Widmer-Cooper, Harrowell and coworkers has shown that key aspects of DH are indeed structural in origin~\cite{CHF04,CH06,H08}.  They do so through the use of the ``isoconfigurational ensemble", a simulation procedure in which a given liquid configuration is analyzed by conducting a set of runs all initiated from the same configuration, but in which particle velocities are randomized~\cite{CHF04}.  Analysis of the ``dynamic propensity", a particle's displacement averaged over all the runs of the IC ensemble, reveals spatial heterogeneity that can only be due to structural properties of the initial configuration, because the influence of the initial velocities has been averaged out.  Simultaneously, there has been significant progress in identifying exactly which local structural properties (e.g. potential energy, soft vibrational modes, medium-range order) may be correlated to DH in several simulated liquids~\cite{FMV06,MRP06,H08,CSK08, tanaka}.

There have also been important recent advances in our understanding of the length scales associated with DH.  Several recent works have quantified the length scale of DH as found in computer simulations of the Kob-Andersen (KA) liquid~\cite{KA1}, an 80:20 binary mixture of A and B Lennard-Jones particles which has received much attention in simulations of glass-forming liquids in general~\cite{SDS98}, and in the analysis of dynamical heterogeneity in particular~\cite{P1,P2,D99,ARMK06,B1,B2,BJ07,SA08,sastry}.  A number of studies have estimated the characteristic length scale $\xi$ for DH from the behavior of the four-point structure factor $S_4(q,t)$ (the Fourier transform of a four-point density correlation function) at small wavenumber $q$~\cite{B1,B2,BJ07,SA08,sastry}.  These studies confirm that $\xi$ increases as $T$ decreases, and yield estimates of $\xi$ ranging from $1$ to $5$ particle diameters in the $T$ region accessible to simulations.  At the same time, these studies emphasize the challenges associated with finding $\xi$ from the small-$q$ behavior of $S_4(q,t)$, due to the limitations of system size.

Sastry and coworkers have recently evaluated $\xi$ for DH in the KA liquid both from a finite-size scaling analysis, and from the behavior of $S_4(q,t)$~\cite{sastry}.  At a reduced temperature of $T=0.6$, they found $\xi=2.6$.  However, at the same $T$ they showed that finite-size effects continue to influence the estimate of $\xi$ as obtained from $S_4(q,t)$ up to a system size of approximately $N=350\, 000$ particles, and that even larger system sizes would be required to accurately evaluate $\xi$ from $S_4(q,t)$ at lower $T$.  In a cubic simulation box with sides of length $L$, a system of $N=350\,000$ particles at the density studied ($\rho=1.2$) has a maximum accessible length scale of $L/2\simeq 33$, i.e. more than an order of magnitude larger than $\xi=2.6$.  The difference between $\xi$ and the size of the system required to accurately compute it demonstrates there exist phenomena that influence DH on length scales many times the value of $\xi$.  This highlights the care that must be taken when interpreting the meaning of length scales associated with DH.

In addition, Berthier and Jack have evaluated $\xi$ for the dynamic propensity in the KA liquid as obtained from the four-point structure factor, generalized so as to quantify the spatial correlations of the dynamics in the isoconfigurational ensemble~\cite{BJ07}.  They report values of $\xi$ for the dynamic propensity ranging from $1$ to $2$.  For the same $T$, the values of $\xi$ found using conventional averaging fall in exactly the same range.  The similarity in the values of $\xi$ obtained from conventional and isoconfigurational averaging suggests that these two approaches probe the same fundamental length scale.  The study of Berthier and Jack is the only one of which we are aware that reports $\xi$ as obtained from the dynamic propensity.  Yet their results demonstrate that the dynamic propensity is a relevant measure for improving our understanding of the spatial correlations associated with DH.

In this paper, we present simulation results exploring the nature of the spatial correlations of the dynamic propensity in the KA liquid.  In the investigations summarized above, the four-point structure factor $S_4(q,t)$ plays a central role.  In order to complement and illuminate these investigations, here we focus instead on quantifying the spatial correlations of the dynamic propensity in real space.  We exploit several approaches:  (i) an evaluation of a real-space correlation function; (ii) a cluster-size analysis of a subset of particles with extremal values of the dynamic propensity; and, (iii) 3D visualizations of the spatial variation of the dynamic propensity field.  As shown below, over a wide range of $T$ we find that significant correlations in the dynamic propensity occur at distances larger than might be expected, given the values of $\xi$ obtained previously by analyzing $S_4(q,t)$.  

In addition, we investigate the use of isoconfigurational averaging for quantities other than the particle displacements.  In a previous analysis of simulated water, we found that the isoconfigurational average of a molecule's potential energy correlated well with its dynamic propensity, suggesting a link between average structure and dynamics at the local level~\cite{MRP06}.  Here we apply this approach to the KA liquid, instead focussing on the coordination of the majority A particles by the minority B particles.  Analogous to our results for water, we find a good spatial correlation between the dynamic propensity and the ``coordination propensity".  Surprisingly, we also find that the magnitude of the spatial correlations of the coordination propensity exceed those of the dynamic propensity, with the most dramatic differences occurring at high $T$.  We discuss the implications of this finding for understanding the relevant length scales of the KA liquid.

\section{Simulation Methods}

Our model system is the KA liquid, consisting of an 80:20 mixture of $N=8788$ A and B particles, interacting via a potential $V_{\alpha\beta}=4\epsilon_{\alpha\beta}[(\sigma_{\alpha\beta}/r)^{12}-(\sigma_{\alpha\beta}/r)^6]$ with $\alpha,\beta \in {\{\rm {A,B}\}}$. All quantities are reported here in reduced units, with length, energy and time given relative to $\sigma_{AA}$, $\epsilon_{AA}$ and $(m\sigma^2_{AA}/48\epsilon_{AA})^{1/2}$ respectively, where $m$ is the mass of the particles. The potential parameters are $\sigma_{AA}=1.0$, $\epsilon_{AA}=1.0$, $\sigma_{BB}=0.88$, $\epsilon_{BB}=0.5$, $\sigma_{AB}=0.8$ and $\epsilon_{AB}=1.5$~\cite{KA1}.  The potential is truncated and shifted at a cutoff radius of 2.5. All simulations are conducted in a cubic cell having sides of length $L$ with periodic boundary conditions and volume fixed to give a density of $1.2$.  The simulation time step is $\Delta t=0.01$.  In all cases below, we restrict our attention to the properties of the A particles.

We study the liquid using the IC ensemble method at $T=1.0$, $0.6$, $0.5$ and $0.466$.  At each $T$ we generate 10 independent starting configurations.  We first equilibrate a random configuration at $T=5.0$ for at least $28\,000$ time steps, then reset $T$ to the desired value, controlling $T$ throughout with a Berendsen thermostat.  Each system is equilibrated for at least $20\tau_\alpha$ where $\tau_\alpha$ is the value of the $\alpha$-relaxation time at that $T$.  We then use each starting configuration to initiate $M=500$ runs of an IC ensemble by randomizing the velocities of all particles according to a Maxwell-Boltzmann distribution, while leaving the particle coordinates unchanged.  The IC ensemble runs are carried out in the microcanonical ensemble.  

Let $r^2(i,k,t)$ be the squared displacement of the $i$-th A particle at time $t$ in run $k$ of an IC ensemble.  The ``dynamic propensity" of each A particle at time $t$ is defined in Ref.~\cite{CHF04} as the the value of, 
\begin{equation}
\langle r_i^2 \rangle_{\rm ic} = \frac{1}{M}\sum_{k=1}^M r^2(i,k,t).  
\end{equation}
We evaluate $\langle r_i^2 \rangle_{\rm ic}$ for each A particle as a function of $t$.

For reference, we show in Fig.~\ref{ngp} the mean squared displacement in the IC ensemble,
\begin{equation} 
\langle r^2 \rangle = \frac{1}{N_AM}\sum_{i=1}^{N_A}\sum_{k=1}^M r^2(i,k,t), 
\end{equation}
and the non-Gaussian parameter, 
\begin{equation} 
\alpha=\frac{3\langle r^4\rangle}{5 \langle r^2\rangle^2}-1,
\end{equation}
where $N_A$ is the number of A particles.  Both quantities show the characteristic pattern of a glass-forming liquid in which DH occurs~\cite{D99}.  $\langle r^2 \rangle$
develops a plateau at low $T$ indicating the onset of molecular
caging, and $\alpha$ displays an increasingly prominent maximum as $T$
decreases. 

Motivated by the results of Ref.~\cite{MRP06}, we also analyze a structural property of the system using the same IC averaging employed to find the dynamic propensity.  Specifically, we examine the chemical composition of the nearest neighbor environment of the A particles.  Let $n(i,k,t)$ be the number of B particles found within a distance of $r=1.2$ (the first minimum of the A-B radial distribution function) of the $i$-th A particle at time $t$ in run $k$ of an IC ensemble.  We define the ``coordination propensity" of each A particle at time $t$ as,
\begin{equation}
\langle n_i \rangle_{\rm ic} = \frac{1}{M}\sum_{k=1}^M n(i,k,t).  
\end{equation}

\section{Spatial Correlation Functions}

We define the spatial correlation function of the propensity as follows.  At a given time $t$, each A particle $i$ has associated with it a value of $\langle r_i^2 \rangle_{\rm ic}$ and $\langle n_i \rangle_{\rm ic}$.  
If we let $x(i,t)=\langle r_i^2 \rangle_{\rm ic}$ or $\langle n_i \rangle_{\rm ic}$, then a spatial correlation function $C(r,t)$ for either propensity can be specified via the following definitions:

\begin{eqnarray}
\langle x(t) \rangle&=& \frac{1}{N_A}\sum_{i=1}^{N_A} x(i,t)\\
\delta x(i,t)&=&x(i,t)-\langle x(t)\rangle\\
\langle [\delta x(t)]^2 \rangle&=& \frac{1}{N_A}\sum_{i=1}^{N_A} [\delta x(i,t)]^2\\
C(r,t)&=&\frac{\langle \delta x(i,t) \delta x(j,t)\rangle}{\langle [\delta x(t)]^2\rangle}
\end{eqnarray}
where $\langle \delta x(i,t) \delta x(j,t)\rangle$ is the average of the product $\delta x(i,t) \delta x(j,t)$ for all pairs of A particles $i$ and $j$ such that $|{\bf r}_i - {\bf r}_j|$ falls within an interval of width $\Delta r$ centered on $r$.  We use $\Delta r=0.1$.  Note that the position of each particle is its position in the starting configuration of the chosen IC run.  All the results for $C(r,t)$ presented below are averaged over the 10 independently-initialized IC runs.  So defined, $C(r,t)$ measures the average spatial correlation in the fluctuation of the propensity (evaluated at time $t$) from its mean value, for particles separated by a distance $r$ in the initial configuration.  If we use $x(i,t)=\langle r_i^2 \rangle_{\rm ic}$, we denote the correlation function for the dynamic propensity as $C_d(r,t)$.  If $x(i,t)=\langle n_i \rangle_{\rm ic}$, then the correlation function for the coordination propensity is denoted $C_c(r,t)$.

\begin{figure}
\centerline{\includegraphics[scale=0.45]{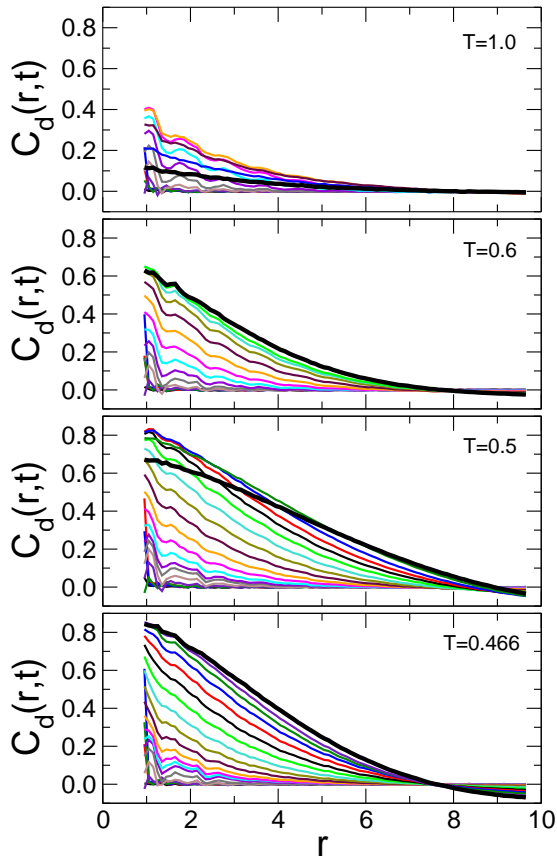}}
\caption{Spatial correlation function of the dynamic propensity.  At each $T$, we show the variation of $C_d(r,t)$ with $r$ for each $t$ studied.  The $t$ values of the curves correspond to the times of the data points in Fig.~\protect{\ref{ngp}}.  The curves for the smallest times decay rapidly to zero.  At intermediate times, the magnitude of the correlation for a given value of $r$ reaches a maximum.  The curve for the largest $t$ is plotted as a thick black line.}
\label{T-100}
\end{figure}

\begin{figure}
\centerline{\includegraphics[scale=0.45]{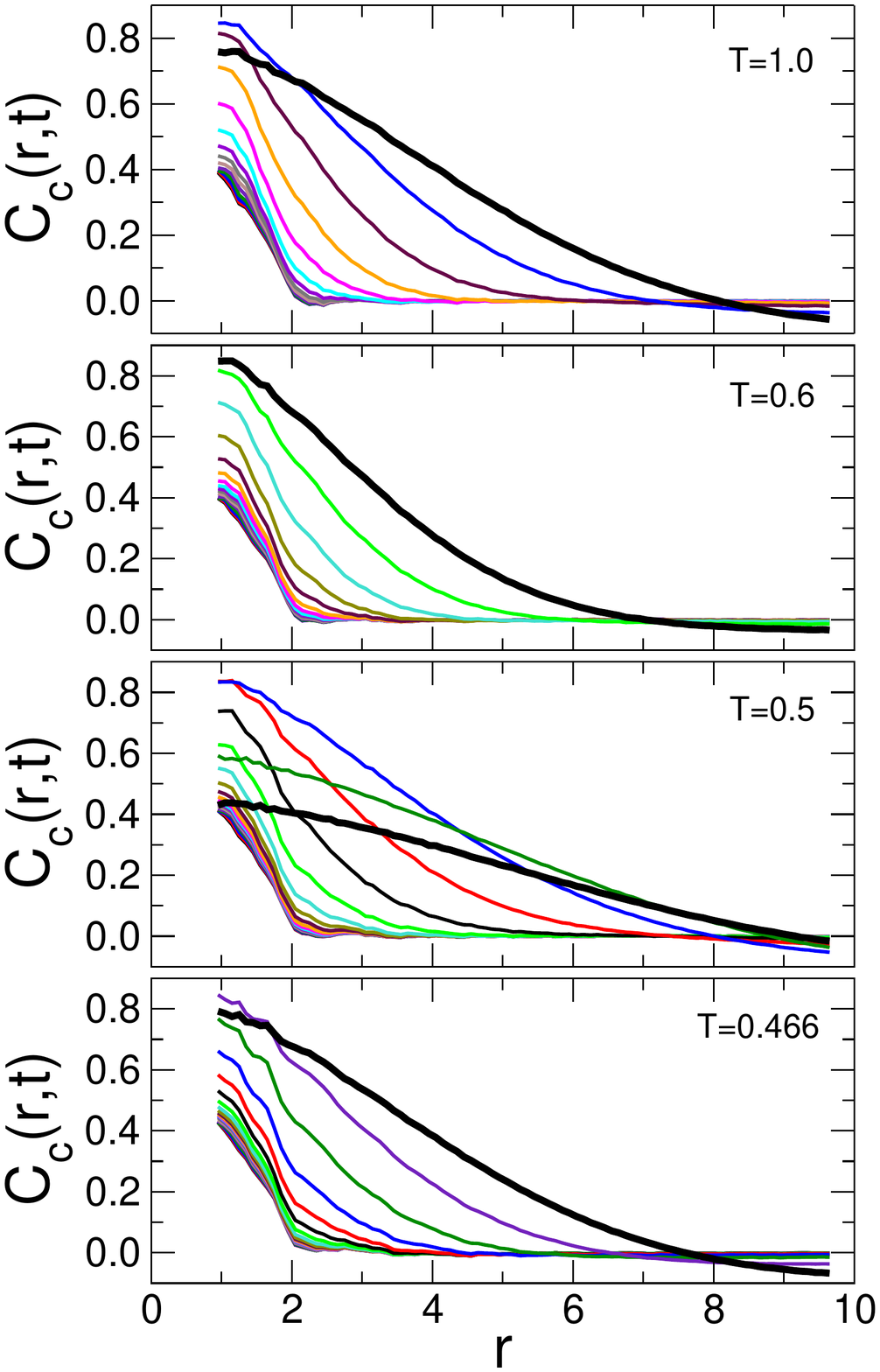}}
\caption{Spatial correlation function of the coordination propensity.  At each $T$, we show the variation of $C_c(r,t)$ with $r$ for each $t$ studied.  The $t$ values of the curves correspond to the times of the data points in Fig.~\protect{\ref{ngp}}.  The curves for the smallest times decay rapidly to zero.  At intermediate times, the magnitude of the correlation for a given value of $r$ reaches a maximum.  The curve for the largest $t$ is plotted as a thick black line.}
\label{T-200}
\end{figure}

\begin{figure}
\centerline{\includegraphics[scale=0.45]{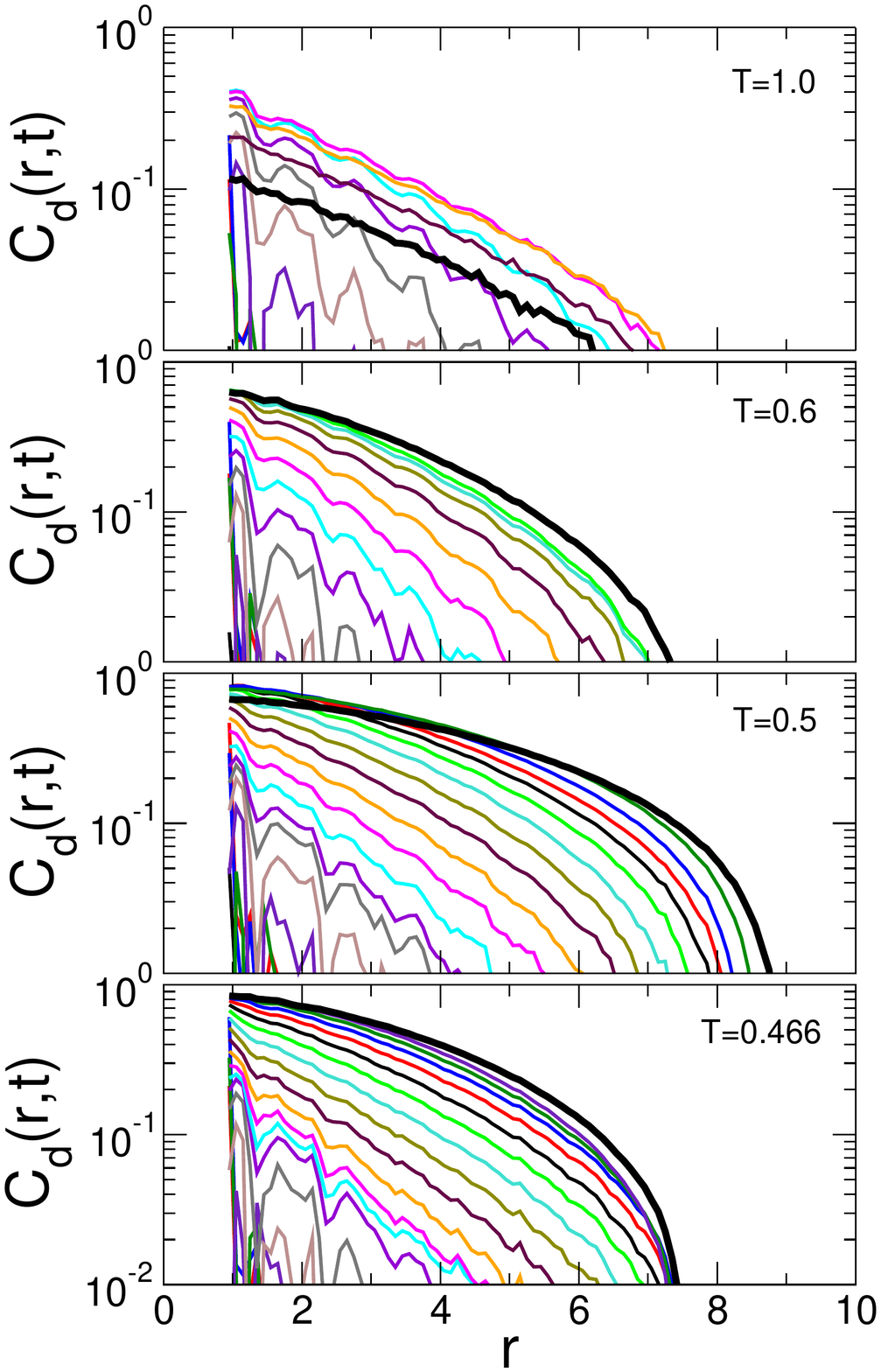}}
\caption{Same data as in Fig.~\protect{\ref{T-100}}, but presented on a semi-log plot.}
\label{T-100-log}
\end{figure}

\begin{figure}
\centerline{\includegraphics[scale=0.45]{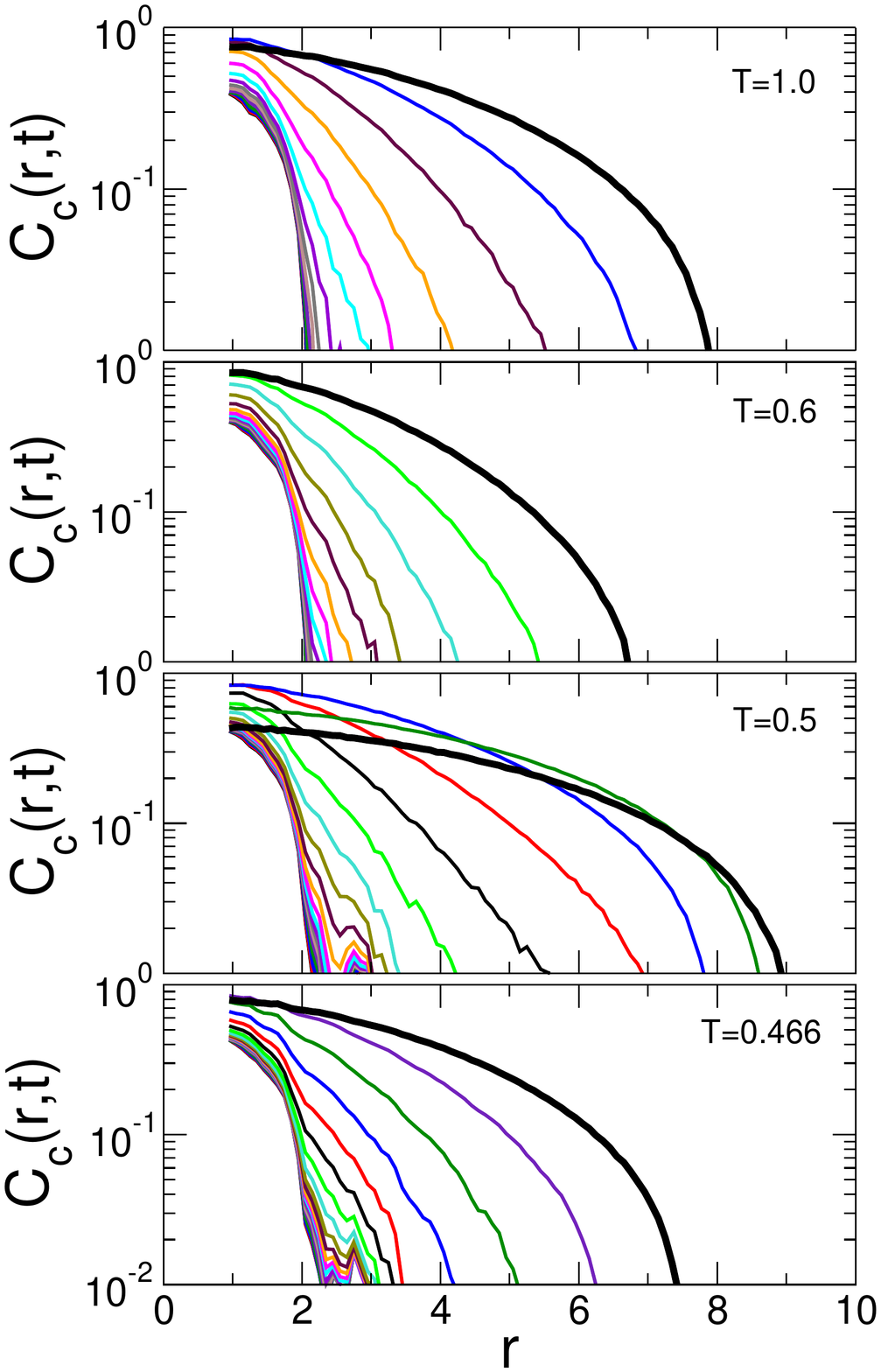}}
\caption{Same data as in Fig.~\protect{\ref{T-200}}, but presented on a semi-log plot.}
\label{T-200-log}
\end{figure}

Figs.~\ref{T-100} and \ref{T-200} respectively show the behavior of $C_d(r,t)$ and $C_c(r,t)$ as a function of both $r$ and $t$, for all four $T$ studied here.  Figs.~\ref{T-100-log} and \ref{T-200-log} show the same data in semi-log form.  

\begin{figure}
\centerline{\includegraphics[scale=0.325]{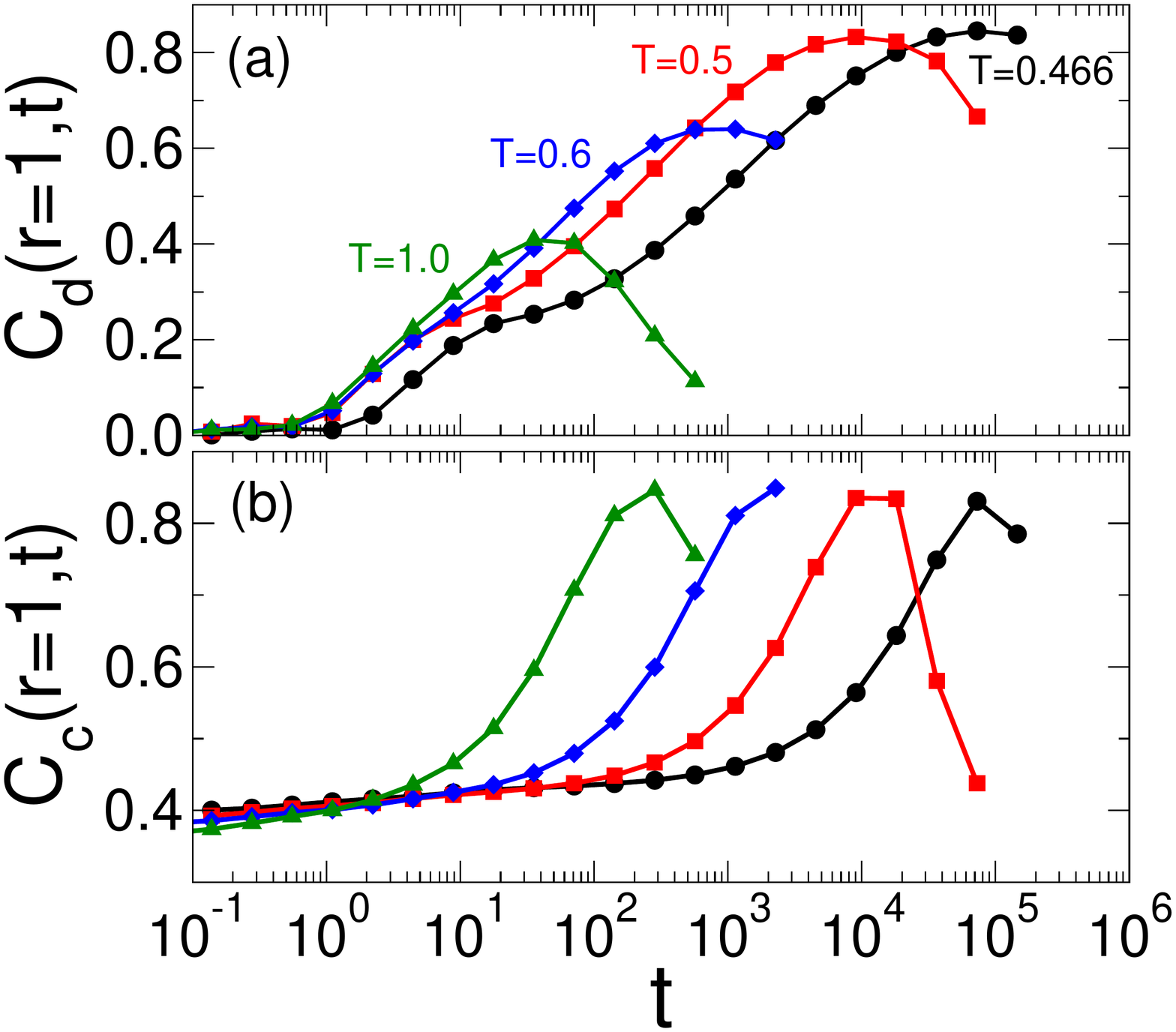}}
\caption{Magnitude of the correlation evaluated at $r=1$, as a function of $t$, for (a) the dynamic propensity, and (b) the coordination propensity.}
\label{corr-amp}
\end{figure}

Focussing first on the the behavior of $C_d(r,t)$, at all $T$ we find that the $r$ dependence of the correlation decays quickly to zero for small values of $t$.  However, as $t$ increases to values corresponding to the vicinity of the maximum in $\alpha$, both the magnitude and range of the correlation increases, reflecting the increasing prominence of DH on this time scale.  At the longest times studied here, the magnitude of the correlation for a given value of $r$ passes through a maximum and begins to decrease.  This effect is most apparent in our data at $T=1.0$ and $0.5$.  We illustrate this behavior in Fig.~\ref{corr-amp}, where we show the value of $C_d(r,t)$ at $r=1.0$, the position of the first maximum in the A-A radial distribution function.   We note that the maxima of the curves in Fig.~\ref{corr-amp} occur more than an order of magnitude later in time than the corresponing maxima of $\alpha$ shown in Fig.~\ref{ngp}.  

The qualitative pattern of behavior we find for $C_d(r,t)$ is what we would expect based on previous work~\cite{P1,P2,D99,ARMK06,B1,B2,BJ07,SA08,sastry}:  Similar to conventional measures of DH, spatial correlations of the dynamic propensity are most pronounced on an intermediate time scale, and both this time scale, and the maximum strength of the correlations, grow as $T$ decreases.  However, two quantitative aspects of the behavior of $C_d(r,t)$ are noteworthy.  First, as shown in Fig.~\ref{T-100-log}, while $C_d(r,t)$ appears to decay exponentially with $r$ for small $t$, at larger $t$ corresponding to the maximum in the magnitude of the correlations, the $r$ dependence is distinctly non-exponential.  Second, at the largest $r$ accessible in our system (i.e. approaching $r=L/2=9.7$), negative correlations occur, and become more prominent as $T$ decreases.  We note that an error analysis over our 10 independent starting configurations confirms that $C_d(r,t)=0$ lies outside the statistical error bars of the correlations at large $r$ and $t$. The functional form of the $r$ dependence of $C_d(r,t)$ for large $t$ is therefore not simple, e.g. exponential decay.   The behavior depicted in Fig.~\ref{T-100} also shows that for $T\leq 0.6$, $C_d(r,t)$ has not reached its asymptotic limit for large $r$ on the scale of the system studied here.  Indeed, the occurrence of negative correlations suggests that  $C_d(r,t)$ may have the form of a damped oscillation extending out to much larger distances than those probed in our system.

Turning next to the behavior of $C_c(r,t)$, we find a similar pattern of behavior as that found for $C_d(r,t)$.  The magnitude and range of the correlations initially increase as $t$ increases (Fig.~\ref{T-200}).  The magnitude of the correlation (at a given $r$) then passes through a maximum and decreases for large $t$ [Fig.~\ref{corr-amp}(b)].  We also find that the $r$ dependence of $C_c(r,t)$ is increasingly non-exponential as $t$ increases (Fig.~\ref{T-200-log}), and that negative correlations occur at the largest values of $r$ (Fig.~\ref{T-200}).

However, there are two prominent differences in the behavior of $C_c(r,t)$ as compared to $C_d(r,t)$.  First, while $C_d(r,t)$ shows negligible correlations at small $t$, $C_c(r,t)$ reveals that a significant correlation exists in the coordination number of nearby particles even in the limit $t\to 0$ (Fig.~\ref{T-200}).  This corresponds to a conventional static correlation that could be computed from instantaneous snapshots of the system configuration.  However, these static correlations are of rather short range, extending out to approximately the second-neighbour shell before dying out completely.  The second difference in the behavior of $C_c(r,t)$ is that the magnitude and range of the correlations at intermediate time are large at all $T$, including at our highest $T=1.0$.  Indeed, there is very little $T$ dependence in the maximum values of the curves plotted in Fig.~\ref{corr-amp}(b).  At all $T$, the behavior of $C_c(r,t)$ suggests that the $r$ dependence of the correlation has not asymptotically vanished on the scale of the system studied here.

\begin{figure}
\centerline{\includegraphics[scale=0.45]{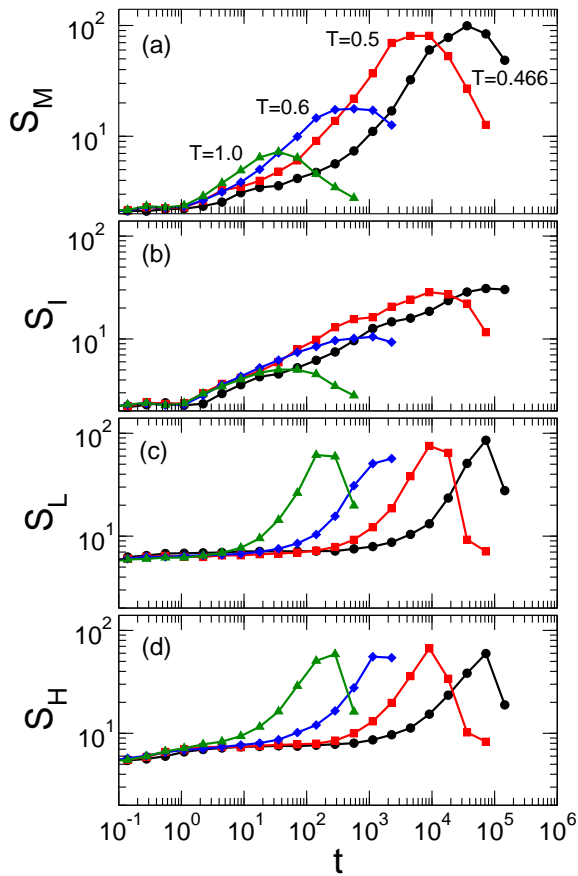}}
\caption{Mean cluster size as a function of $t$ for (a) mobile clusters, (b) immobile clusters, (c) clusters of A particles with low B coordination, and (d) clusters of A particles with high B coordination, all evaluated in the IC ensemble.}
\label{S}
\end{figure}

\begin{figure}
\centerline{\includegraphics[scale=0.3]{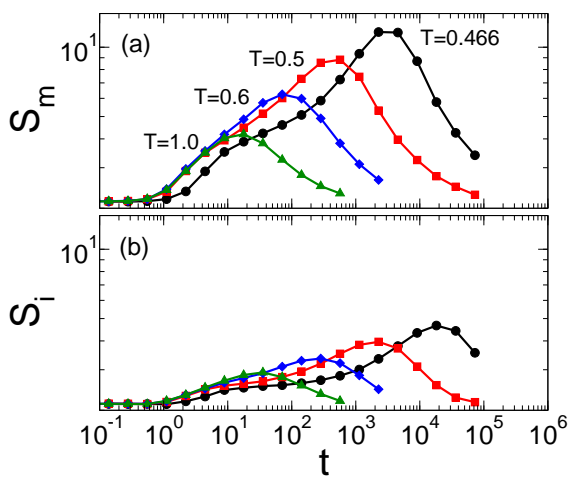}}
\caption{Mean cluster size as a function of $t$ for (a) mobile clusters and (b) immobile clusters, as measured in single simulations runs.}
\label{strings}
\end{figure}

\begin{figure}
\centerline{\includegraphics[scale=0.37]{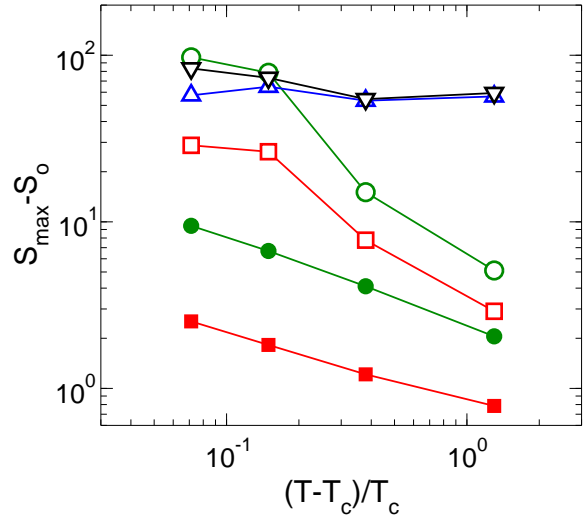}}
\caption{Maximum value $S^{\rm max}$ of the mean cluster size for each of the curves plotted in Figs.~\ref{S} and \ref{strings}, as a function of $(T-T_c)/T$, where $T_c=0.435$.  Shown are the maximum values of $S_i$ (filled squares), $S_m$ (filled circles), $S_I$ (open squares), $S_M$ (open cicles), $S_H$ (up triangles) and $S_L$ (down triangles).  $S_o=2.135$ has been subtracted from all the data to reflect the value of $S^{\rm max}$  in excess of its value for a random choice of $10\%$ of the A particles.  For the open symbols, the standard deviation of the data over the 10 starting configurations is less than or comparable to the symbol size.}
\label{max}
\end{figure}

The non-trivial $r$ dependence of the correlation functions shown in Figs.~\ref{T-100} and \ref{T-200} (including the possibility of negative correlations) does not lend itself to a straight-forward quantification of a characteristic length scale from either $C_d(r,t)$ or $C_c(r,t)$.  Indeed, it seems that systems much larger than that studied here would be required to determine the full range of $r$ over which significant non-zero correlations occur.  This observation is consistent with the much larger system sizes found in Ref.~\cite{sastry} to be required for the accurate evaluation of $\xi$.

We note that given the faster-than-exponential decay of the correlation functions shown in Figs.~\ref{T-100-log} and \ref{T-200-log}, it may be reasonable to fit the curves using a ``compressed" exponential function, 
\begin{equation}
C_{\rm fit}(r)=C_o\exp[-(r/\xi)^\zeta]-C_1,
\label{cexp}
\end{equation}
where $\zeta>1$.  Eq.~\ref{cexp} was found to fit the data for the decay of the ``overlap" correlations of a soft-sphere binary mixture studied in Ref.~\cite{cavagna}.  While the fits to our data (not shown) in many cases are quite good, this is only the case when we include the shift term $C_1$ as a fitting parameter, which is required in order to account for the negative correlations appearing at large $r$.  However, since the correlation functions must approach zero as $r\to \infty$, this fitting form cannot be a complete description of the $r$ dependence of the correlations, and we do not pursue this further here.

\section{Cluster-Size Analysis}

As an alternative to evaluating a characteristic length scale from our spatial correlation functions, we explore the typical size of the correlations using a cluster-size analysis of a subset of particles having extremal values of the dynamic propensity.  This approach was widely used in earlier work on DH, and while more qualitative in nature, succeeded in capturing many of the key trends for how the size of DH correlations grow as $T$ decreases~\cite{D99,VG04,MRP06}.  

To this end, we identify the particles having the highest $10\%$ of $\langle r_i^2\rangle_{\rm ic}$ values at a given $t$, and then find the clusters of ``mobile particles" formed by this subset.  Clusters are defined by the criterion that two particles of the subset that are also within $r=1.4$ of one another (the position of the first minimum of the A-A radial distribution function) in the initial configuration are assigned to the same cluster.  The number-averaged mean cluster size of a set of $N_c$ clusters is,
\begin{equation}
S=\frac{1}{N_c}\sum_nn{\cal N}(n),
\end{equation}
where ${\cal N}(n)$ is the number of clusters of size $n$.  We evaluate $S$ for the clusters of mobile particles defined above, and denote it as $S_M$.  We conduct the same analysis on the lowest $10\%$ of $\langle r_i^2\rangle_{\rm ic}$ values, and find the mean cluster size of this ``immobile" subset $S_I$.  Figs.~\ref{S}(a) and (b) show the $t$ dependence of $S_M$ and $S_I$, where the data are averaged over the 10 starting configurations used at each $T$.

For comparison, we also evaluate $S$ for the DH that occurs in single simulation runs.  That is, we separately analyze each of the $M=500$ runs of one IC ensemble at each $T$, and evaluate, as a function of $t$, the mean cluster size $S_m$ of the clusters formed by the particles having the largest $10\%$ of displacements as measured from their position in the starting configuration. These mobile clusters correspond to the ``strings" documented e.g. in Ref.~\cite{D99}.  We then obtain the average of the $S_m$ curves over all $500$ runs [Fig.~\ref{strings}(a)].  The corresponding cluster-size analysis of the smallest $10\%$ of displacements in individual runs gives $S_i$ as a function of $t$ [Fig.~\ref{strings}(b)].  

Fig.~\ref{S}(a,b) and Fig.~\ref{strings} allow us to compare the DH as revealed by both IC and conventional averaging, for both mobile and immobile domains.  The $t$ dependence of all curves follows the behavior for DH found in earlier work (see e.g. Ref.~\cite{VG04}).  At small $t$, $S$ has the value expected for a random choice of $10\%$ of the A particles (approximately $S_o=2.135$), consistent with no spatial correlations.  However, on the time scale of structural relaxation a maximum occurs, indicating significant clustering of mobile and immobile particles.  At large $t$, the DH begins to dissipate and $S$ decreases toward $S_o$.  This pattern of behavior is entirely consistent with that found for the correlation function $C_d(r,t)$.  Also, the $t$ dependence of $S_M$ and $S_I$ is quite similar to that found for the magnitude of $C_d(r,t)$ for $r=1$ depicted in Fig.~\ref{corr-amp}.

The monotonic increase in $S^{\rm max}$ (the maximum value of $S$) as $T$ decreases quantifies the growth of DH on cooling.  We denote the maximum value of $S_M$ in Fig.~\ref{S}(a) as $S^{\rm max}_M$; similarly, $S^{\rm max}_I$, $S^{\rm max}_m$ and $S^{\rm max}_i$ denote the maxima in Figs.~\ref{S}(b), \ref{strings}(a) and \ref {strings}(b) respectively.
The $T$ dependence of $S^{\rm max}$ for all data in Figs.~\ref{S} and \ref{strings} is plotted in Fig.~\ref{max} as a function of $(T-T_c)/T_c$, where $T_c=0.435$ is the critical temperature of mode coupling theory for the KA liquid.  Fig.~\ref{max} shows that the sizes of the mobile and immobile clusters found using the IC ensemble are as much as an order of magnitude larger than the DH found when analyzing single runs.  Also, the $T$ dependence of $S^{\rm max}$ is quite different for the two kinds of averaging.  $S^{\rm max}_m$ and $S^{\rm max}_i$ follow a power law~\cite{D99,FN1}, while $S^{\rm max}_M$ and $S^{\rm max}_I$ do not.  Indeed, the most notable behavior in Fig.~\ref{max} is that $S^{\rm max}_M$ and $S^{\rm max}_I$ both initially grow faster than a power law on cooling, but then at the lowest $T$ their growth seems to saturate.  This behavior is consistent with the size of the largest clusters becoming ``capped" by the size of the system at the lowest $T$.  Hence, as in the case of $C_d(r,t)$, our cluster-size analysis suggests that the size of the correlated domains of the dynamic propensity exceed our system size as $T$ decreases.

We have also carried out the same cluster-size analysis on the coordination propensity.  We show in Fig.~\ref{S}(c) the time dependence of the mean cluster size $S_L$ of the $10\%$ of A particles with the lowest values of $\langle n_i \rangle_{\rm ic}$.  Fig.~\ref{S}(d) shows $S_H$, the mean cluster size of the $10\%$ of A particles with the highest values of $\langle n_i \rangle_{\rm ic}$.  The time evolution of $S_L$ and $S_H$ in Fig.~\ref{S} is similar to $S_M$ and $S_I$, with the notable exception that the values of $S^{\rm max}_L$ and $S^{\rm max}_H$ (shown in Fig.~\ref{max}) are nearly independent of $T$, and are larger than or comparable to $S^{\rm max}_M$ and $S^{\rm max}_L$ for all $T$.  That is, the size of the domains with high and low B coordination (as quantified by the coordination propensity) remains large at all $T$, including at high $T$ where the mobile and immobile domains (as quantified by the dynamic propensity) are an order of magnitude smaller.   These findings are consistent with the comparison of the $C_d(r,t)$ and $C_c(r,t)$ correlation functions presented in the previous sections.  We also note that the instantaneous spatial correlations in the local coordination of A particles [observed in the $t \to 0$ limit of $C_c(r,t)$] are reflected in the $t \to 0$ limit of $S_L$ and $S_H$ in Fig.~\ref{S}.  In this limit, $S_L$ and $S_H$ approach values that are distinctly greater than the random value $S_o$, reflecting the presence of instantaneous correlations of the local coordination.

\begin{figure*}
\centerline{\includegraphics[scale=0.60]{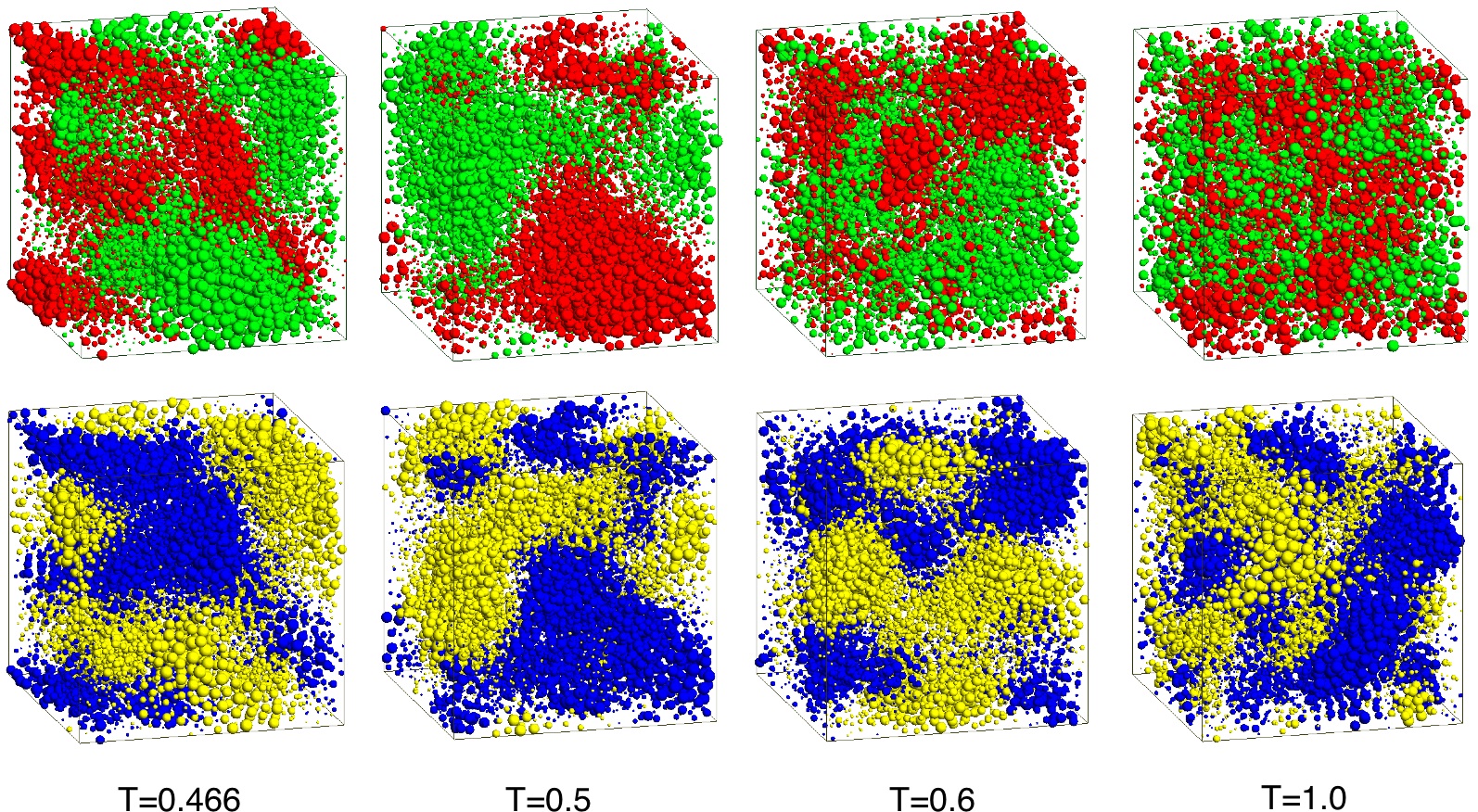}}
\caption{Spatial variation of $\langle r_i^2 \rangle_{\rm ic}$ (top panels) and  $\langle n_i \rangle_{\rm ic}$ (bottom panels) at each $T$.  To make each top panel, the values of $\langle r_i^2 \rangle_{\rm ic}$, evaluated at the time of the maximum of $S_M$, are assigned to each A particle at its position in the initial configuration of the IC ensemble.  These values are sorted and assigned an integer rank $R_i$ from 1 to $N$, from smallest to largest.  Each A particle is then plotted as a green sphere of radius $\sigma=R_{\rm min}\exp\{[(R_i-N)/(1-N)]\log(R_{\rm max}/R_{\rm min})\}$, where $R_{\rm max}=0.5$ and $R_{\rm min}=0.01$.  The ranks $R_i$ are then reversed (i.e. assigned from largest to smallest), and each A particle is also plotted as a red sphere of radius $\sigma$.  The color observed for each particle therefore indicates which of the green or red spheres is larger.  The result presents the rank of $\langle r_i^2 \rangle_{\rm ic}$ on an exponential scale, such that the largest green spheres represent the most mobile A particles, and the largest red spheres the most immobile.  The bottom panels are created in exactly the same way as the top panels, but with $\langle r_i^2 \rangle_{\rm ic}$ replaced by $\langle n_i \rangle_{\rm ic}$, and where the time is chosen to be the maximum of $S_L$ at each $T$.  In the bottom panels, the largest yellow spheres represent the A particles with the lowest B coordination, and the largest blue spheres the A particles with the highest B coordination.}
\label{pics}
\end{figure*}

\section{Visualizing the Dynamic and Coordination Propensity}

In the top panels of Fig.~\ref{pics} we visualize the spatial variation of the dynamic propensity for one starting configuration at each $T$, at the value of $t$ corresponding to $S^{\rm max}_M$, where the clusters are most prominent.  Our procedure is based on that used in Ref.~\cite{MRP06}.  Particles in the top (bottom) $50\%$ of $\langle r_i^2 \rangle_{\rm ic}$ values are represented as green (red) spheres, with each sphere plotted at the position of the particle in the initial configuration.  The radius of each sphere represents the rank order of $\langle r_i^2 \rangle_{\rm ic}$: the larger a green (red) sphere is, the larger (smaller) is its value of $\langle r_i^2 \rangle_{\rm ic}$.  See the caption of Fig.~\ref{pics} for complete details of the visualization procedure.

Note that these visualizations represent all the A particles in the system, not just those in the top or bottom 10\% of $\langle r_i^2 \rangle_{\rm ic}$ values that are used in the previous section to define clusters and obtain the mean cluster size.  Nonetheless, the pattern of heterogeneity observed in Fig.~\ref{pics} is entirely consistent with the cluster-size analysis presented in Fig.~\ref{S}, and with the behaviour of the spatial correlation functions in Fig.~\ref{T-100}.  At $T=1.0$ the arrangement of red and green spheres is nearly random, while at the lowest $T$, very large mobile and immobile domains have emerged.  At both $T=0.5$ and $0.466$, the domains are strikingly large, and are comparable to the system size.  This is consistent with the possibility that finite-size effects are responsible for the saturation of the values of  $S^{\rm max}_M$ and $S^{\rm max}_I$ observed at low $T$ in Fig.~\ref{max}.

The spatial variation of the coordination propensity $\langle n_i \rangle_{\rm ic}$ is visualized in the bottom panels of Fig.~\ref{pics} for the same initial configurations shown in the top panels, with $t$ chosen at the time of $S^{\rm max}_L$.  In all other respects, the coordination propensity values are represented in the same way as the dynamic propensity values, except that the particles in the top (bottom) $50\%$ of $\langle n_i \rangle_{\rm ic}$ values are represented as blue (yellow) spheres, instead of green (red).  These visualizations confirm that the size of the domains with low and high B coordination remain large at all $T$, including at high $T$ where the mobile and immobile domains are much smaller.  

Further, Fig.~\ref{pics} illustrates that the locations of the mobile and immobile domains that emerge on cooling approximately correspond with the domains of low and high B coordination that are prominent at all $T$.  This spatial correspondence suggests a correlation between average local dynamics and average local structure that would be consistent with expectation:  Given the attractive interaction between A and B particles, an A particle with lower-than-average B coordination will be less tightly bound by its neighbors, and thus potentially more mobile, than one with higher-than-average B coordination.  

\section{Discussion}

Our main result is to present a quantification of the correlations of the dynamic propensity as they occur in real space.  As shown above, these correlations continue to have a significant $r$ dependence on the largest length scales accessible in our system of $N=8788$ particles.  Previous simulation studies~\cite{P1,P2,D99,ARMK06,B1,B2,BJ07} of the KA liquid that address DH typically consider systems for which $N$ ranges from $1000$ to $8000$ particles, and so our system is comparable to the largest systems in this range.  The notable exceptions are the recent works that study systems of $N=27\,000$~\cite{SA08} and $N=350\,000$~\cite{sastry} particles.  

Our work shows that a definitive study of the correlations of the dynamic propensity should be undertaken in a system much larger than that used here.  This includes studies conducted at high $T$, i.e. up to twice $T_c$ or even higher.  If it is true that the length scale of the dynamic propensity and the length scale of DH as found in conventional averaging are the same, then our results support the finding of Ref.~\cite{sastry} that systems very much larger than ours are required in order to study DH in a regime that is beyond the influence of finite-size effects, even at high $T$.  The same conclusion is even more strongly supported by the behavior of the coordination propensity;  these correlations span our system size at all $T$ studied, including $T=1.0$.

We emphasize that since our main results are based on isoconfigurational averaging, the structural triggers for individual correlated dynamical events occurring in a single simulation run (e.g. the ``strings'' of Ref.~\cite{D99}) are not specifically addressed here.  Isoconfigurational averaging quantifies the {\it tendency} for a particular property (a displacement or a coordination number) to be observed, but has very limited predictive power for any given run~\cite{BJ07}. 

It is also important to recognize that the correlations exposed via isoconfigurational averaging are properties of the initial configuration from which the runs of the isoconfigurational ensemble are generated.  That is, they are {\it static} correlations in the sense that they are a property determined by the configuration of particles in an instantaneous snapshot of the system~\cite{CHF04}.  At the same time, the role of the subsequent dynamical evolution of the ensemble of systems initiated from the original configuration in revealing these correlations cannot be ignored.  It is the sensitivity of the dynamics to the initial configuration that reveals the spatial heterogeneity observed in the propensity.  In short, isoconfigurational averaging provides us with a {\it dynamically-revealed} static correlation, but a static correlation nonetheless.  In this light, the large-scale correlations (that is, large compared to our system size) that we observe in the dynamic propensity, and especially in the coordination propensity at $T=1.0$, indicate the existence of subtle configurational fluctuations in the KA liquid that are larger than has perhaps been generally appreciated.  The nature of these structural fluctuations clearly merits further study, for example, in terms of fluctuations of local composition or medium-range order that occur without accompanying density fluctuations, as examined in the recent work of Tanaka and coworkers~\cite{tanaka}.

In all glass-forming liquids, the increasing sensitivity of dynamics to structure as $T$ decreases makes it inevitable that local dynamical fluctuations (i.e. DH) will occur on a scale at least up to the size of any local structural fluctuations that are present.  The spatial extent of the fluctuations we find in the coordination propensity are comparable to our system size at all $T$ studied, and as described above, these fluctuations are necessarily structural in origin.  Our results therefore suggest that the occurrence of DH in the KA liquid can be understood as a response, progressively emerging as $T$ decreases, of the local dynamics to subtle but large-scale structural fluctuations that are already well-established at high $T$.  

\acknowledgments
We thank ACEnet for providing computational resources, and NSERC, CFI and AIF for financial support.  GSM is supported by an ACEnet Research Fellowship, and PHP by the CRC program.


\begin{thebibliography}{999}
\bibitem{FN1} While power law growth of the size of mobile clusters in single runs is well known, our results show that the immobile clusters also exhibit a power law growth with approximately the same exponent as for mobile clusters.
\end{thebibliography}
\end{document}